\documentclass{article}
\include{input}

\usepackage{spconf,amsmath,graphicx,mathtools,algorithm2e}
\PassOptionsToPackage{hyphens}{url}\usepackage{hyperref}
\usepackage{multirow}
\usepackage{enumitem}

\hypersetup{colorlinks=true}
\usepackage{todonotes}

\title{Fully supervised speaker diarization}
\name{Aonan Zhang\textsuperscript{1,2} \quad Quan Wang\textsuperscript{1} \quad Zhenyao Zhu\textsuperscript{1} \quad John Paisley\textsuperscript{2} \quad Chong Wang\textsuperscript{1}
\thanks{The first author performed this work as an intern at Google. The implementation of the algorithms in this paper is available at: \mbox{\url{https://github.com/google/uis-rnn}}}}
\address{\textsuperscript{1}Google Inc., USA \qquad \textsuperscript{2}Columbia University, USA\\[4pt] {
  \normalsize
  \textsuperscript{1}
   \{
    \href{mailto:aonan@google.com}{\nolinkurl{aonan}},
    \href{mailto:quanw@google.com}{\nolinkurl{quanw}},
    \href{mailto:zyzhu@google.com}{\nolinkurl{zyzhu}},
    \href{mailto:chongw@google.com}{\nolinkurl{chongw}}
    \}
  {\tt @google.com} \qquad
  \textsuperscript{2}
   \{
    \href{mailto:az2385@columbia.edu}{\nolinkurl{az2385}},
    \href{mailto:jpaisley@columbia.edu}{\nolinkurl{jpaisley}}
    \}
  {\tt @columbia.edu}
}}

\begin{document}
\ninept
\maketitle
\begin{abstract}
\label{sec:abstract}
In this paper, we propose a fully supervised speaker diarization approach, named {\it unbounded interleaved-state recurrent neural networks} (UIS-RNN). Given extracted speaker-discriminative embeddings (\textit{a.k.a.} d-vectors) from input utterances, each individual speaker is modeled by a parameter-sharing RNN, while the RNN states for different speakers interleave in the time domain. This RNN is naturally integrated with a distance-dependent Chinese restaurant process (ddCRP) to accommodate an unknown number of speakers. Our system is fully supervised and is able to learn from examples where time-stamped speaker labels are annotated. We achieved a 7.6\% diarization error rate on NIST SRE 2000 CALLHOME, which is better than the state-of-the-art method using spectral clustering. Moreover, our method decodes in an online fashion while most state-of-the-art systems rely on offline clustering.
\end{abstract}
\section{Introduction}
\label{sec:intro}

Aiming to solve the problem of ``who spoke when'', most existing speaker diarization systems consist of multiple relatively independent components \cite{sell2015diarization,garcia2017speaker,wang2017speaker}, including but not limited to: (1) A speech segmentation module, which removes the non-speech parts, and divides the input utterance into small segments; (2) An embedding extraction module, where speaker-discriminative embeddings such as speaker factors \cite{castaldo2008stream}, i-vectors \cite{dehak2011front}, or d-vectors \cite{ge2e} are extracted from the small segments; (3) A clustering module, which determines the number of speakers, and assigns speaker identities to each segment; (4) A resegmentation module, which further refines the diarization results by enforcing additional constraints \cite{sell2015diarization}.

For the embedding extraction module, recent work \cite{garcia2017speaker,wang2017speaker,zajic2017speaker} has shown that the diarization performance can be significantly improved by replacing i-vectors \cite{dehak2011front} with neural network embeddings, \textit{a.k.a.} d-vectors \cite{ge2e,heigold2016end}. This is largely due to the fact that neural networks can be trained with big datasets, such that the model is sufficiently robust against varying speaker accents and acoustic conditions in different use scenarios. 

However, there is still one component that is unsupervised in most modern speaker diarization systems --- the clustering module. Examples of clustering algorithms that have been used in diarization systems include Gaussian mixture models \cite{zajic2017speaker,shum2013unsupervised}, mean shift \cite{senoussaoui2014study},  agglomerative hierarchical clustering \cite{garcia2017speaker,sell2014speaker}, k-means \cite{wang2017speaker,dimitriadis2017developing}, Links \cite{wang2017speaker,mansfield2018links}, and spectral clustering \cite{wang2017speaker,ning2006spectral}.


Since both the number of speakers and the segment-wise speaker labels are determined by the clustering module, the quality of the clustering algorithm is critically important to the final diarization performance. However, the fact that most clustering algorithms are unsupervised means that, we will not able to improve this module by learning from examples when the time-stamped speaker labels ground truth are available. In fact, in many domain-specific applications, it is relatively easy to obtain such high quality annotated data.

In this paper, we replace the unsupervised clustering module by an online generative process that naturally incorporates labelled data for training. We call this method \textit{unbounded interleaved-state recurrent neural network} (UIS-RNN), based on these facts: (1) Each speaker is modeled by an instance of RNN, and these instances share the same parameters; 
(2) An unbounded number of RNN instances can be generated; (3) The states of different RNN instances, corresponding to different speakers, are interleaved in the time domain.
Within a fully supervised framework, our method in addition handles complexities in speaker diarization: it automatically learns the number of speakers within each utterance via a Bayesian non-parametric process, and it carries information through time via the RNN.

The contributions of our work are summarized as follows:
\begin{enumerate}[noitemsep]
    \item Unbounded interleaved-state RNN, a trainable model for the general problem of segmenting and clustering temporal data by learning from examples.
    \item Framework for a fully supervised speaker diarization system.
    \item New state-of-the-art performance on NIST SRE 2000 CALLHOME benchmark.
    \item Online diarization solution with offline quality.
\end{enumerate}

\section{Baseline system using clustering}
\label{sec:baseline}

Our diarization system is built on top of the recent work by Wang \textit{et al.} \cite{wang2017speaker}. Specifically, we use exactly the same segmentation module and embedding extraction module as their system, while replacing their clustering module by an unbounded interleaved-state RNN.

\begin{figure}
  \centering\vspace{-5pt}
    \includegraphics[width=0.45\textwidth]{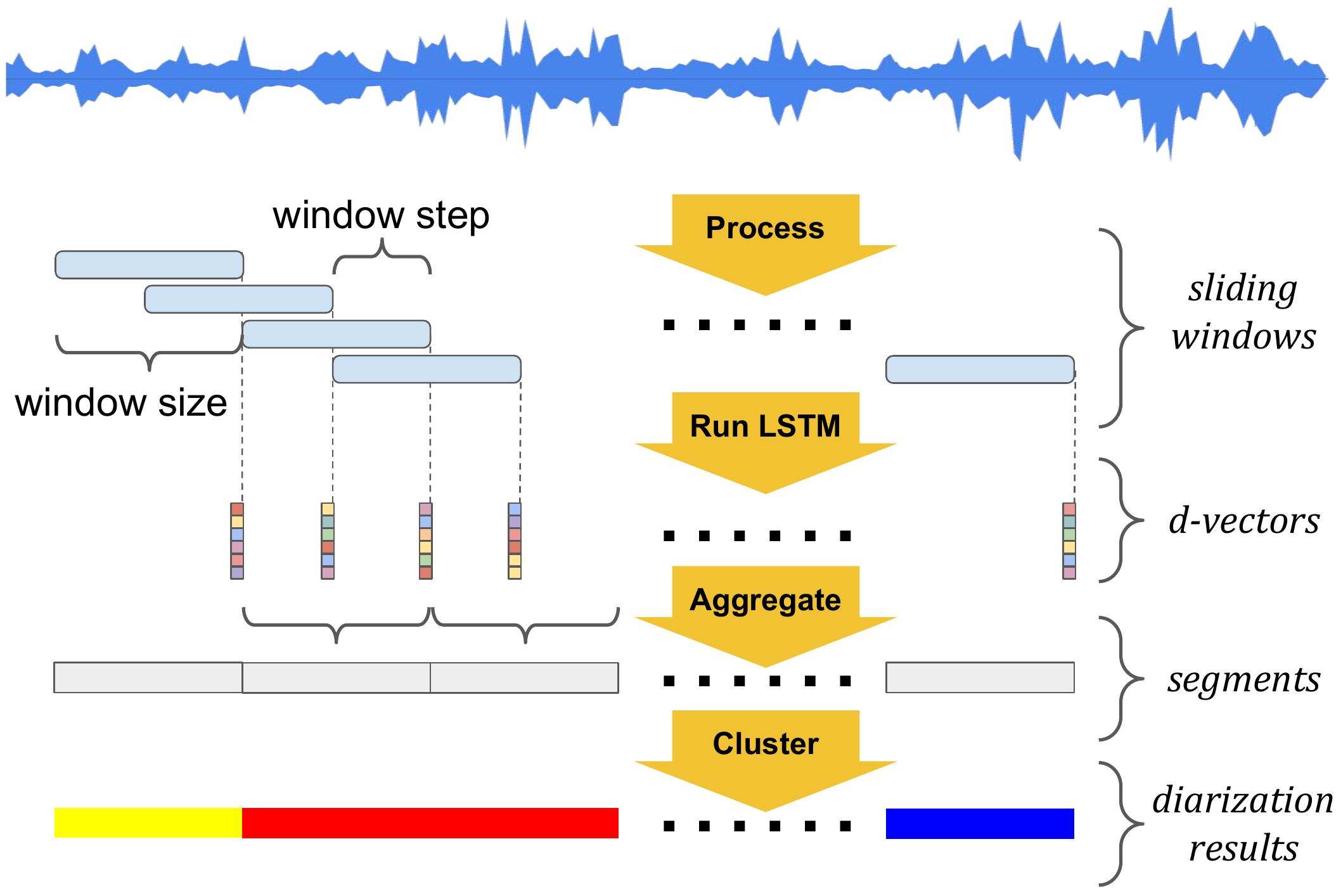}\vspace{-10pt}
  \caption{The baseline system architecture \cite{wang2017speaker}.}\vspace{-5pt}
  \label{fig:baseline}
\end{figure}

As a brief review, in the baseline system \cite{wang2017speaker}, a text-independent speaker recognition network is used to extract embeddings from sliding windows of size 240ms and 50\% overlap. A simple voice activity detector (VAD) with only two full-covariance Gaussians is used to remove non-speech parts, and partition the utterance into non-overlapping segments with max length of 400ms. Then we average window-level embeddings to segment-level d-vectors, and feed them into the clustering algorithm to produce final diarization results. The workflow of this baseline system is shown in Fig. \ref{fig:baseline}.

The text-independent speaker recognition network for computing embeddings has three LSTM layers and one linear layer. The network is trained with the state-of-the-art generalized end-to-end loss \cite{ge2e}. We have been retraining this model for better performance, which will be later discussed in Section~\ref{sec:dvector}.

\section{Unbounded interleaved-state RNN}
\label{sec:model}

\subsection{Overview of approach}

Given an utterance, from the embedding extraction module, we get an \textit{observation sequence} of embeddings $\bfX=(\bfx_1,\bfx_2,\ldots,\bfx_T)$, where each $\bfx_t\in\bbR^d$.
Each entry in this sequence is a real-valued d-vector corresponding to a segment in the original utterance.
In the supervised speaker diarization scenario, we also have the ground truth speaker labels for each segment $\bfY=(y_1,y_2,\ldots,y_T)$.
Without loss of generality, let $\bfY$ be a sequence of positive integers by the order of appearance.

For example, $\bfY=(1,1,2,3,2,2)$ means this utterance has six segments, from  three different speakers, where $y_t=k$ means segment $t$ belongs to speaker $k$.

UIS-RNN is an online generative process of an entire utterance $(\bfX,\bfY)$, where\footnote{We denote an ordered set $(1,2,\ldots,t)$ as $[t]$.}
\begin{align}
    p(\bfX,\bfY)=p(\bfx_1,y_1)\cdot\prod_{t=2}^T p(\bfx_t,y_t|\bfx_{[t-1]},y_{[t-1]}).
\end{align}
To model speaker changes, we use an augmented representation
\begin{align}
    \hspace{-4pt}p(\bfX,\bfY,\bfZ)\!=\!p(\bfx_1,y_1)\!\cdot\!\prod_{t=2}^T p(\bfx_t,y_t,z_t|\bfx_{[t-1]},y_{[t-1]},z_{[t-1]}),
    \label{eq:augmented_joint_probability}
\end{align}
where $\bfZ=(z_2,\ldots,z_T)$, and $z_t=\mathbbm{1}(y_t\neq y_{t-1})\in\{0,1\}$ is a binary indicator for speaker changes. For example, if $\bfY=(1,1,2,3,2,2)$, then $\bfZ=(0,1,1,1,0)$. Note that $\bfZ$ is uniquely determined by $\bfY$, but $\bfY$ cannot be uniquely determined by a given $\bfZ$, since we don't know which speaker we are changing to. Here we leave $z_1$ undefined, and factorize each product term in Eq.~(\ref{eq:augmented_joint_probability}) as three parts that separately model sequence generation, speaker assignment, and speaker change:
\begin{align}
    &p(\bfx_t,y_t,z_t|\bfx_{[t-1]},y_{[t-1]},z_{[t-1]})\nonumber\\
    &\quad=\underbrace{p(\bfx_t|\bfx_{[t-1]},y_{[t]})}_{\text{sequence generation}}\cdot\underbrace{p(y_t|z_t,y_{[t-1]})}_{\text{speaker assignment}}\cdot\underbrace{p(z_t|z_{[t-1]})}_{\text{speaker change}}.
\end{align}
For the first entry of the sequence, we let $y_1=1$ and there is no need to model speaker assignment and speaker change. In Section~\ref{subsec:submodels}, we introduce these components separately.

\subsection{Details on model components}
\label{subsec:submodels}

\subsubsection{Speaker change}
\label{subsec:speaker_change}

We assume the probability of $z_t \in \{0,1\}$ follows:
\begin{align}
p(z_t=0|z_{[t-1]},\bflambda)=g_{\bflambda}(z_{[t-1]}),
    \label{eq:generic_generate_z}
\end{align}
where $g_{\bflambda}(\cdot)$ is a function paramaterized by $\bflambda$. Since $z_t$ indicates speaker change at time $t$, we have
\begin{align}
    p(y_t=y_{t-1}|z_t,y_{[t-1]})~=~1-z_t.
    \label{eq:generate_y_1}
\end{align}

In general, $g_{\bflambda}(\cdot)$ could be any function, such as an RNN. But for simplicy, in this work, we make it a constant value $g_{\bflambda}(z_{[t-1]})=p_0 \in [0,1]$.
This means $\{z_t\}_{t \in [2,T]}$ are independent binary variables parameterized by $\bflambda=\{p_0\}$:
\begin{align}
    z_t\sim_{iid.}\text{Binary}(p_0).
    \label{eq:generate_z}
\end{align}

\subsubsection{Speaker assignment process}

One of the biggest challenges in speaker diarization is to determine the total number of speakers for each utterance. To model the speaker turn behavior in an utterance, we use a distance dependent Chinese restaurant process (ddCRP)~\cite{blei2011distance}, a Bayesian non-parametric model that can potentially model an unbounded number of speakers.
Specifically, when $z_t=0$, the speaker remains unchanged. When $z_t=1$, we let
\begin{align}
    p(y_t=k|z_t=1,y_{[t-1]})~&\propto~N_{k,t-1},\nonumber\\
    p(y_t=K_{t-1}+1|z_t=1,y_{[t-1]})~&\propto~\alpha.
    \label{eq:generate_y_2}
\end{align}
Here $K_{t-1}\defeq\max y_{[t-1]}$ is the total number of unique speakers up to the $(t-1)$-th entry. Since $z_t=1$ indicates a speaker change, we have $k\in[K_{t-1}]\setminus\{y_{t-1}\}$. In addition, we let $N_{k,t-1}$ be the number of \textit{blocks} for speaker $k$ in $y_{[t-1]}$. A \textit{block} is defined as a maximum-length subsequence of \textit{continuous} segments that belongs to a single speaker. For example, if $y_{[6]}=(1,1,2,3,2,2)$, then there are four blocks $(1,1)|(2)|(3)|(2,2)$ separated by the vertical bar, with $N_{1,5}=1,N_{2,5}=2,N_{3,5}=1$.
The probability of switching back to a previously appeared speaker is proportional to the number of continuous speeches she/he has spoken. There is also a chance to switch to a new speaker, with a probability proportional to a constant $\alpha$. The joint distribution of $\bfY$ given $\bfZ$ is
\begin{align}
    \hspace{-5pt}p(\bfY|\bfZ,\alpha)=\cfrac{\alpha^{K_T-1}\prod_{k=1}^{K_T}\Gamma(N_{k,T})}{\prod_{t=2}^T (\sum_{k\in [K_{t-1}]\setminus\{y_{t-1}\}}N_{k,t-1}+\alpha)^{\mathbbm{1}(z_t=1)}}.
    \label{eq:generate_y_3}
\end{align}

\subsubsection{Sequence generation}

Our basic assumption is that, the observation sequence of speaker embeddings $\bfX$ is generated by distributions that are parameterized by the output of an RNN. This RNN has multiple instantiations, corresponding to different speakers, and they share the same set of RNN parameters $\bftheta$. In our work, we use gated recurrent unit (GRU)~\cite{cho2014learning} as our RNN model, to memorize long-term dependencies.

At time $t$, we define $\bfh_t$ as the state of the GRU corresponding to speaker $y_t$,
and 
\begin{align}
    \bfm_t=f(\bfh_t | \bftheta)
    \label{eq:m_t}
\end{align}
as the output of the entire network, which may contain other layers.
Let $t'\defeq\max\{0,~s<t:~y_s=y_{t}\}$ be the last time we saw speaker $y_{t}$ before $t$, then:
\begin{align}
\bfh_{t}=\text{GRU}(\bfx_{t'},\bfh_{t'}|\bftheta),
\end{align}
where we can assume $\bfx_0=\bf0$ and $\bfh_0=\bf0$, meaning all GRU instances are initialized with the same zero state.

Based on the GRU outputs, we assume the speaker embeddings are modeled by:\vspace{-5pt}
\begin{align}
    \bfx_t|\bfx_{[t-1]},y_{[t]}\sim\cN(\bfmu_t,\sigma^2 \bfI),
    \label{eq:generate_x}
\end{align}
where $\bfmu_t=(\sum_{s=1}^t \mathbbm{1}(y_{s}=y_t))^{-1}\cdot(\sum_{s=1}^t \mathbbm{1}(y_{s}=y_t)\bfm_{s})$ is the averaged GRU output for speaker $y_t$.

\begin{figure}
  \centering
    \includegraphics[width=0.44\textwidth]{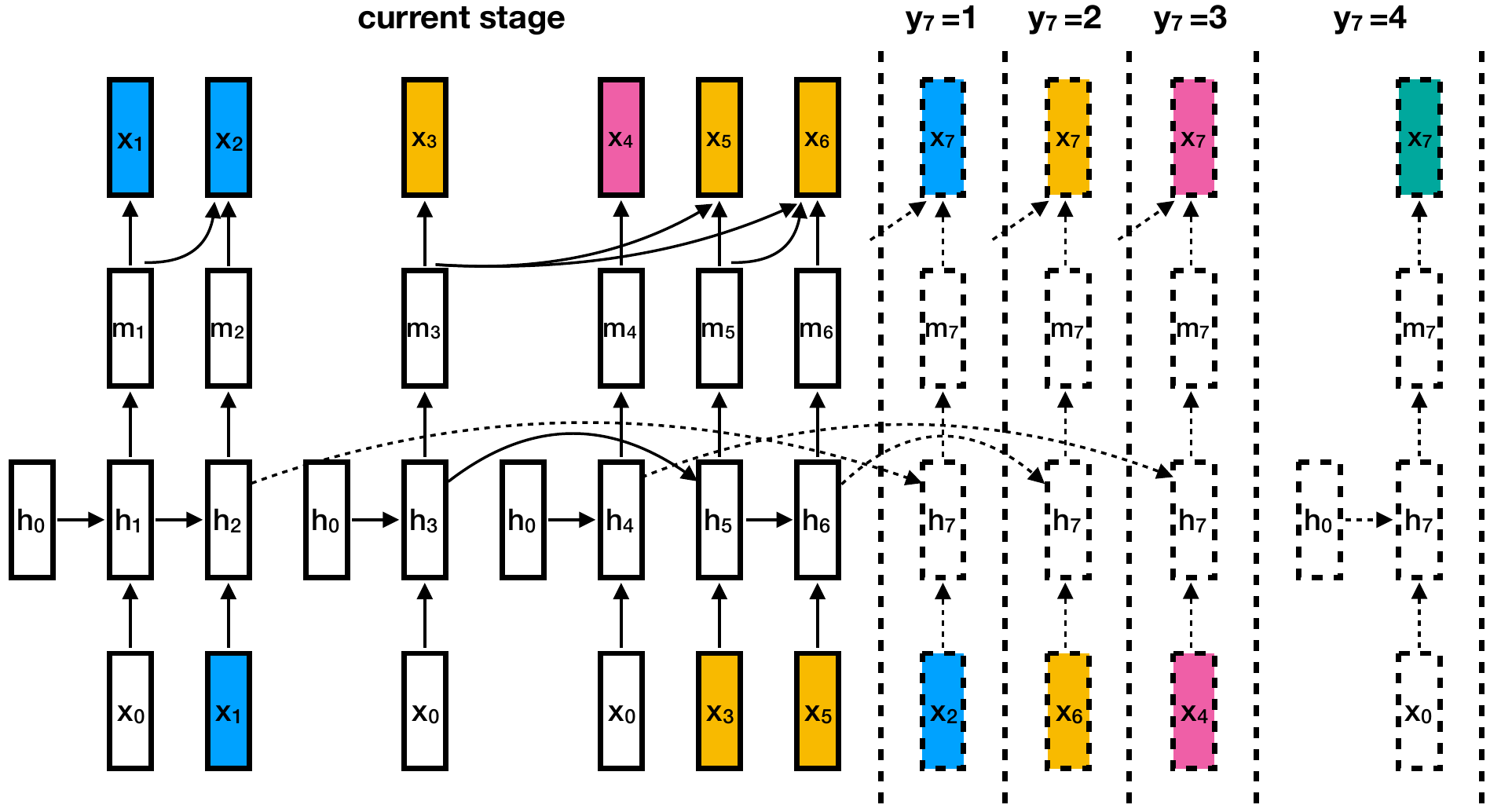}\vspace{-10pt}
  \caption{Generative process of UIS-RNN. Colors indicate labels for speaker segments. There are four options for $y_7$ given $\bfx_{[6]},y_{[6]}$.}\vspace{-10pt}
  \label{fig:full_model}
\end{figure}

\subsubsection{Summary of the model}

We briefly summarize UIS-RNN in Fig.~\ref{fig:full_model}, where $\bfZ$ and $\bflambda$ are omitted for a simple demonstration. At the current stage (shown in solid lines) $y_{[6]}=(1,1,2,3,2,2)$. There are four options for $y_7$: $1,2,3$ (existing speakers), and $4$ (a new speaker). The probability for generating a new observation $\bfx_7$ (shown in dashed lines) depends both on previous label assignment sequence $y_{[6]}$, and previous observation sequence $\bfx_{[6]}$.

\subsection{MLE Estimation}
Given a training set $(\bfX_1,\bfX_2,\ldots,\bfX_N)$ containing $N$ utterances together with their labels $(\bfY_1,\bfY_2,\ldots,\bfY_N)$, we maximize the following log joint likelihood:\vspace{-5pt}
\begin{align}
    \max_{\bftheta,\alpha,\sigma^2,\bflambda}\quad\sum_{n=1}^N\ln p(\bfX_n,\bfY_n,\bfZ_n|~\bftheta,\alpha,\sigma^2,\bflambda).
    \label{eq:train_joint_likelihood}
\end{align}

Here we include all hyper-parameters, and each term in Eq.~(\ref{eq:train_joint_likelihood}) can be factorized exactly as Eq.~(\ref{eq:augmented_joint_probability}). 

The estimation of $\bflambda$ depends on how $g_{\bflambda}(\cdot)$ is defined. When we simply have $g_{\bflambda}(z_{[t-1]})=p_0$,
we have a closed-form solution:
\begin{align}
    p_0^*=\frac{\sum_{n=1}^N\sum_{t=2}^{T_n}\mathbbm{1}(y_{n,t}=y_{n,t-1})}{\sum_{n=1}^N T_n-N},
\end{align}
where $T_n$ denotes the sequence length of the $n$th utterance.

For $\bftheta$ and $\sigma^2$, there is no closed-form update. We use stochastic gradient ascent by randomly selecting a subset $B^{(\tau)}\subset[N]$ of $|B^{(\tau)}|=b$ utterances. For $\bftheta$, we update:
\begin{align}
    \bftheta^{(\tau)}\!=\!\bftheta^{(\tau-1)}\!+\!\frac{N\rho^{(\tau)}}{b}\sum_{n\in B^{(\tau)}}\nabla_{\bftheta} \ln p(\bfX_n|~\bfY_n,\bfZ_n,\bftheta,-),
    \label{eq:update_theta}
\end{align}
since $\bftheta$ is independent of $(\bfY_n,\bfZ_n)$. Eq.~(\ref{eq:update_theta}) also applies to $\sigma^2$ by replacing $\bftheta$ with $\sigma^2$. For $\alpha$, we update
\begin{align}
    \alpha^{(\tau)}\!=\!\alpha^{(\tau-1)}\!+\!\frac{N\rho^{(\tau)}}{b}\sum_{n\in B^{(\tau)}}\nabla_{\alpha} \ln p(\bfY_n|~\bfZ_n,\alpha,-),
    \label{eq:update_theta}
\end{align}
where $p(\bfY_n|~\bfZ_n,\alpha,-)$ is given in Eq.~(\ref{eq:generate_y_3}). In our experiments, we run multiple iterations with a constant step size $\rho^{(\tau)}=\rho$ until convergence. 

\begin{algorithm}[t]
 \KwData{$\bfX^{test}=(\bfx^{test}_1,\bfx^{test}_2,\ldots,\bfx^{test}_T)$}
 \KwResult{$\bfY^*=(y^*_1,y^*_2,\ldots,y^*_T)$}
 initialize $\bfx_0=\bf0,\bfh_0=\bf0$\;
 \For{$t=1,2,\ldots,T$}{
  $(y_t^*,z_t^*)=\mathop{\arg\max}_{(y_t, z_t)} \big( \ln p(z_t)$\hfill {Eq. (\ref{eq:generate_z})}\\
  \qquad\quad$+~\ln p(y_t|z_t,y_{[t-1]}^*)$\hfill {Eq. (\ref{eq:generate_y_1},~\ref{eq:generate_y_2})}\\
  \qquad\quad$+~\ln p(\bfx_t|\bfx_{[t-1]},y_{[t-1]}^*,y_t)\big)$\hfill {Eq. (\ref{eq:generate_x})}\\
 update $N_{k,t-1}$ and GRU hidden states\;
 }
 \caption{Online greedy MAP decoding for UIS-RNN.}
 \label{alg:greedy_decoding}
\end{algorithm}

\subsection{MAP Decoding}
Since we can decode each testing utterance in parallel, here we assume we are given a testing utterance $\bfX^{test}=(\bfx_1,\bfx_2\,\ldots,\bfx_T)$ without labels. The ideal goal is to find 
\begin{align}
    \bfY^*=\mathop{\arg\max}_{\bfY}~\ln p(\bfX^{test},\bfY).
\end{align}\\[-10pt]
However, this requires an exhaustive search over the entire combinatorial label space with complexity $\cO(T!)$, which is impractical. Instead, we use an online decoding approach which sequentially performs a greedy search, as shown in Alg.~\ref{alg:greedy_decoding}. This will significantly reduce computational complexity to $\cO(T^2)$. We observe that in most cases the maximum number of speakers per-utterance is bounded by a constant $C$. In that case, the complexity will further reduce to $\cO(T)$. In practice, we apply a beam search~\cite{medress1977speech} on the decoding algorithm, and adjust the number of look-ahead entries to achieve better decoding results. 

\vspace{-2pt}
\section{Experiments}\vspace{-2pt}
\label{sec:exp}

\subsection{Speaker recognition model}

\label{sec:dvector}

We have been retraining the speaker recognition network with more data and minor tricks (see next few paragraphs) to improve its performance. Let's call the text-independent speaker recognition model in \cite{wang2017speaker,ge2e,jia2018transfer} as \mbox{``d-vector V1''}. This model is trained with 36M utterances from 18K US English speakers, which are all mobile phone data based on anonymized voice query logs.

To train a new version of the model, which we call ``d-vector V2'' \cite{voicefilter}, we added: (1) non-US English speakers; (2) data from far-field devices; (3) public datasets including LibriSpeech \cite{panayotov2015librispeech}, VoxCeleb \cite{nagrani2017voxceleb}, and VoxCeleb2 \cite{chung2018voxceleb2}. The non-public part contains 34M utterances from 138K speakers, while the public part is added to the training process using the MultiReader approach \cite{ge2e}.

Another minor but important trick is that, the speaker recognizer model used in \cite{wang2017speaker} and \cite{ge2e} are trained on windows of size 1600ms, which causes performance degradation when we run inference on smaller windows. For example, in the diarization system, the window size is only 240ms. Thus we have retrained a new model ``d-vector V3'' by using variable-length windows, where the window size is drawn from a uniform distribution within $[240 \textrm{ms}, 1600 \textrm{ms}]$ during training.

The speaker verification Equal Error Rate (EER) of the three models on two testing sets are shown in Table \ref{tab:dvector_eer}. On speaker verification tasks, adding more training data has significantly improved the performance, while using variable-length windows for training also slightly further improved EER.

\begin{table}
\centering
  \caption{Speaker verification EER of the three speaker recognition models. en-ALL represents all English locales. The EER=3.55\% for d-vector V1 on en-US phone data is the same as the number reported in Table 3 of \cite{ge2e}.}
  \label{tab:dvector_eer}
  \begin{tabular}{| c | c | c |}
    \hline
    \multirow{2}{*}{Model} & EER (\%) on en-US  & EER (\%) on en-ALL  \\
    & phone data & phone + farfield data \\ \hline \hline
    d-vector V1 & 3.55 & 6.14 \\ 
    d-vector V2 & 3.06 & 2.03 \\ 
    d-vector V3 & 3.03 & 1.91 \\ \hline
  \end{tabular}\vspace{-15pt}
\end{table}

\vspace{-2pt}
\subsection{UIS-RNN setup}\vspace{-2pt}
\label{subsec:uis-rnn-setup}
For the speaker change, as we have stated in Section~\ref{subsec:speaker_change}, we assume $\{z_t\}_{t \in [2,T]}$ follow independent identical binary distributions for simplicity.

Our sequence generation model is composed of one layer of 512 GRU cells with a tanh activation, followed by two fully-connected layers each with 512 nodes and a ReLU~\cite{nair2010rectified} activation. The two fully-connected layers corresponds to Eq. (\ref{eq:m_t}).

For decoding, we use beam search of width 10.
\vspace{-2pt}
\subsection{Evaluation protocols}\vspace{-2pt}

Our evaluation setup is exactly the same as \cite{wang2017speaker}, which is based on the pyannote.metrics library \cite{bredin2017pyannote}. We follow these common conventions of other works:

\begin{itemize}[noitemsep]
    \item We evaluate on single channel audio.
    \item We exclude overlapped speech from evaluation.
    \item We tolerate errors less than 250ms in segment boundaries. 
    \item We report the confusion error, which is usually directly referred to as Diarization Error Rate (DER) in the literature.\vspace{-7pt}
\end{itemize}

\vspace{-5pt}
\subsection{Datasets}\vspace{-2pt}

For the evaluation, we use 2000 NIST Speaker Recognition Evaluation (LDC2001S97), Disk-8, which is usually directly referred to as ``CALLHOME'' in literature. It contains 500 utterances distributed across six languages: Arabic, English, German, Japanese, Mandarin, and Spanish. Each utterance contains 2 to 7 speakers.

Since our approach is supervised, we perform a 5-fold cross validation on this dataset. We randomly partition the dataset into five subsets, and each time leave one subset for evaluation, and train UIS-RNN on the other four subsets. Then we combine the evaluation on five subsets and report the averaged DER.

Besides, we also tried to use two off-domain datasets for training UIS-RNN: (1) 2000 NIST Speaker Recognition Evaluation, Disk-6, which is often referred to as ``Switchboard''; (2) ICSI Meeting Corpus \cite{janin2003icsi}.
We first tried to train UIS-RNN purely on off-domain datasets, and evaluate on CALLHOME; we then tried to add the off-domain datasets to the training partition of each of the 5-fold.

\vspace{-5pt}
\subsection{Results}\vspace{-2pt}

We report the diarization performance results on 2000 NIST SRE Disk-8 in Table \ref{tab:der_disk8}. For each version of the speaker recognition model, we compare UIS-RNN with two baseline approaches: k-means and spectral offline clustering. For k-means and spectral clustering, the number of speakers is adaptively determined as in~\cite{wang2017speaker}. For UIS-RNN, we show results for three types of evaluation settings: (1) in-domain training (5-fold); (2) off-domain training (Disk-6 + ICSI); and (3) in-domain plus off-domain training.

From the table, we see that the biggest improvement in DER actually comes from upgrading the speaker recognition model from V2 to V3. This is because in V3, we have the window size consistent between training time and diarization inference time, which was a big issue in V1 and V2.

UIS-RNN performs noticeably better than spectral offline clustering, when using the same speaker recognition model. It is also important to note that UIS-RNN inference produces speaker labels in an \textbf{online} fashion. As discussed in \cite{wang2017speaker}, online unsupervised clustering algorithms usually perform significantly worse than offline clustering algorithms such as spectral clustering.

Also, adding more data to train UIS-RNN also improved DER, which is consistent with our expectation -- UIS-RNN benefits from learning from more examples. Specifically, while large scale off-domain training already produces great results in practice (Disk-6 + ICSI), the availability of in-domain data can further improve the performance (5-fold + Disk-6 + ICSI).

\begin{table}
\centering
  \caption{DER on NIST SRE 2000 CALLHOME, with comparison to other systems in literature. VB is short for Variational Bayesian resegmentation \cite{sell2015diarization}. The DER=12.0\% for d-vector V1 and spectral clustering is the same as the number reported in Table 2 of \cite{wang2017speaker}.
  }
  \label{tab:der_disk8}
  \begin{tabular}{| c | c | c | c |}
    \hline
    \bf d-vector & \bf Method & \bf Training data & \bf DER (\%) \\ \hline \hline
    \multirow{5}{*}{V1} & k-means & --- & 17.4  \\
    & spectral & --- & 12.0  \\
    & UIS-RNN &  5-fold & {\bf 11.7}  \\
    & UIS-RNN &  Disk-6 + ICSI & {\bf 11.7}  \\
    & UIS-RNN & 5-fold + Disk-6 + ICSI & {\bf 10.6}\\ \hline
    \multirow{5}{*}{V2} & k-means & --- & 19.1  \\
    & spectral & --- & 11.6  \\
    & UIS-RNN &  5-fold & {\bf 10.9}  \\
    & UIS-RNN &  Disk-6 + ICSI & {\bf 10.8}  \\
    & UIS-RNN & 5-fold + Disk-6 + ICSI & {\bf 9.6}  \\ \hline
    \multirow{5}{*}{V3} & k-means & --- & 12.3  \\
    & spectral & --- & 8.8  \\
    & UIS-RNN &  5-fold & {\bf 8.5}  \\
    & UIS-RNN &  Disk-6 + ICSI & {\bf 8.2}  \\
    & UIS-RNN & 5-fold + Disk-6 + ICSI & {\bf 7.6}  \\ \hline \hline
    \multicolumn{3}{|c|}{Castaldo \textit{et al.} \cite{castaldo2008stream}} & 13.7 \\
    \multicolumn{3}{|c|}{Shum \textit{et al.} \cite{shum2013unsupervised}} & 14.5  \\
    \multicolumn{3}{|c|}{Senoussaoui \textit{et al.} \cite{senoussaoui2014study}} & 12.1  \\
    \multicolumn{3}{|c|}{Sell \textit{et al.} \cite{sell2015diarization} (+VB)} & 13.7 (11.5)  \\
    \multicolumn{3}{|c|}{Garcia-Romero \textit{et al.} \cite{garcia2017speaker} (+VB)} & 12.8 (9.9)  \\
    \hline
  \end{tabular}\vspace{-15pt}
\end{table}





\vspace{-5pt}
\section{Conclusions}\vspace{-5pt}
\label{sec:conclusions}
In this paper, we presented a speaker diarization system where the commonly used clustering module is replaced by a trainable \mbox{unbounded} interleaved-state RNN. Since all components of this \mbox{system} can be learned in a supervised manner, it is preferred over unsupervised systems in scenarios where training data with high quality time-stamped speaker labels are available. On the NIST SRE 2000 CALLHOME benchmark, using exactly the same speaker embeddings, this new approach, which is an online algorithm, outperforms the state-of-the-art spectral offline clustering algorithm.

Besides, the proposed UIS-RNN is a generic solution to the sequential clustering problem, with other potential applications such as face clustering in videos. One interesting future work direction is to directly use accoustic features instead of pre-trained embeddings as the observation sequence for UIS-RNN, such that the entire speaker diarization system becomes an end-to-end model.

\newpage
\bibliographystyle{IEEEbib}
\bibliography{refs}

\end{document}